\documentstyle[12pt,epsfig]{article} 
 \def\baselinestretch{1.3}
\newcommand{\ba}{\begin{array}}
\newcommand{\ea}{\end{array}}
\newcommand{\bd}{\begin{displaymath}}
\newcommand{\ed}{\end{displaymath}}
\newcommand{\be}{\begin{equation}}
\newcommand{\ee}{\end{equation}}
\newcommand{\bea}{\begin{eqnarray}}
\newcommand{\eea}{\end{eqnarray}}

\def\barr{\begin{array}}
\def\earr{\end{array}}

\newcommand{\beqn}{\begin{eqnarray}}
\newcommand{\eeqn}{\end{eqnarray}}

\def\etal{ {\em et al.}~}

\def\q2 {q^2}

\def\go{\rightarrow}

\begin{document}
\begin{titlepage}

\begin{flushright}
{\large OITS-725}
\end{flushright}

\begin{center}
{\Large\bf Higgs pair production in association 
with a vector boson at $e^+e^-$ colliders in theories of 
higher dimensional gravity}\\[10mm]
N.~G.~Deshpande{\footnote{email: desh@oregon.uoregon.edu}}, 
Dilip Kumar Ghosh{\footnote{email: dghosh@physics.uoregon.edu}} 
\\[4mm]
{\em Institute of Theoretical Science \\
     5203 University of Oregon\\
Eugene OR 97403-5203} \\[10mm]
\end{center}
\begin{abstract}
The models of large extra compact dimensions, as suggested by Arkani-Hamed,
Dimopoulos and Dvali, predict exciting phenomenological consequences with 
gravitational interactions becoming strong at the TeV scale. Such theories
can be tested at the existing and future colliders. In this paper, we study 
the contribution of virtual Kaluza-Klein excitations (both spin-0 and 2) 
in the process $e^+e^- \to HHZ$ and $ e^+e^- \to HH\gamma $ 
at future linear collider (NLC). We find that the virtual exchange KK 
gravitons can modify
the cross-section $\sigma(e^+e^- \to HHZ)$ significantly from its Standard 
Model value and will allow the effective string scale to be probed up to 
6.6 TeV. The second process is absent at the tree level 
in the standard model, and can therefore be used to put limits 
on the effective string scale of 7.4 TeV.
 \end{abstract}
 
\end{titlepage}

\vskip 1 true cm

\newpage

\def\baselinestretch{1.8}


\section{Introduction}
\label{sect:intro}

The concept of large extra dimensions and TeV scale  
gravity introduced by Arkhani-Hamed, Dimopoulos and Dvali, better 
known as ADD model~\cite{ADD} has attracted a lot of attention. In this 
scenario, the total space-time has $D= 4+n $ dimensions, where 
gravity lives in $n$ additional large spatial dimensions of size $R$.
The standard model particles are confined to the usual $3+1$-dimension. 

The effective $4D$ Planck 
scale, $M_{Pl}\sim 2.4\times 10^{18}$~GeV , is related to the {\it 
fundamental } Planck scale $M_{S}$ in $(4+n)$ dimension by
\beqn
M^2_{Pl}\sim R^{n} M^{n +2}_{S}
\eeqn 

Thus, for the large extra-dimensions, it is possible to have a 
{\it fundamental} scale $M_{S}$ as low as a TeV~\cite{ADD}, solving the
gauge hierarchy problem of the standard model.  With $M_{S}\sim $ TeV, 
for $n= 1$  the value of $R$ ($\sim 10^{11}$ millimeter ) is
ruled out by gravitational experiments. On the other hand, $n\geq 2 $, we 
get $R\leq $ mm, a range that is still allowed by gravitational experiments. 

Each graviton couples to the standard model matter and gauge particles through 
energy-momentum tensor and its trace, with the strength suppressed by 
powers of the $4$-dimensional Planck scale, $M_{Pl}$. However, from the 
4-dimensional point of view, the massless gravitons propagating 
in the $(4+n)$-dimensional bulk are seen as massive towers of Kaluza-Klein
(KK) modes of excitations with spin-0, spin-1 (which decouples ) 
and spin-2. 
The mass spectrum of these KK modes can be treated as a continuum, 
since their mass splitting $\sim 1/R$ is about $10^{-4}$ eV for 
$n=2$ and for $n=6$ it is of the order of a MeV. After summing over 
these KK modes one gets 
enhancement of KK graviton coupling to standard model 
matter and gauge particles to 
powers of $(\sqrt{s}/M_S)$~\cite{ADD}, where $\sqrt{s}$ is the available
energy for the process. The Feynman rules for this theory have been
developed considering a linearized theory of gravity in the bulk in Ref.
~\cite{GRW, Han-Lykken}. These new interactions can give rise to
several interesting phenomenological consequences testable at present and 
future colliders~\cite{ADD_rev}. The effect of these new interactions can be
observed either through production of real KK modes, or through the exchange
of virtual KK modes in various processes~\cite{ADD_rev, TESLA, ANLC}. 

In this paper we will consider the process $e^+e^- \to HHZ$ and 
$e^+e^- \to HH\gamma$ to study the effect of low-scale gravity at 
proposed linear colliders with center of mass energy 500 GeV and beyond. 
However, one should keep in mind that these processes will not serve as
the dominant discovery channel for KK gravitons, since there are more direct 
processes for its discovery \cite{ADD_rev}. Once the discovery of KK 
gravitons is established, then the next phase will be to look for their 
effect on some more complicated processes, like the ones discussed here, to
re-confirm their discovery.  

In the standard model, $HHZ$ production has been studied in the context of the
determination of the Higgs self couplings ~\cite{zerwas}. 
The other channel, $HH\gamma$, is a unique process mediated by virtual 
gravitons. This process is absent at the tree level 
in the standard model, therefore, the observation
of such a process can lead to a possible indication of low scale 
gravity. Through out this paper we assume that the Higgs boson will already
have been discovered and its mass determined.    

The rest of the paper is organized as follows: In Section 2, we study the
production of $HHZ$ and discuss the effect of KK gravitons. In Section 3,
we will discuss how virtual exchange of gravitons give rise to $HH\gamma$
final state and discuss the consequences. Section 4 is reserved for the overall
discussions and conclusions. In Appendix, we present the graviton exchange 
amplitudes for $e^+ e^- \to HHZ$ process. 

\section{$e^+e^- \to HHZ $ }
In this section we study the additional contribution of virtual exchange of 
spin 2 and 0 graviton modes in models of large extra dimensions 
to the Higgs pair production in association with $Z$.
In the standard model, the process $e^+ e^- \to HHZ $ has been studied 
as a means of determining $HHH$ couplings. The standard 
model amplitudes and polarized cross-sections are given in the 
Ref.\cite{zerwas}. In our
analysis, we shall use unpolarized beams. Additional diagrams arising from 
the exchange of both spin-2 and spin-0 KK gravitons are shown 
in Figure 1, where $KK$ stands for both the spin-2 and spin-0 modes.
The amplitudes using the Feynman rules \footnote{In our analysis
we take $\xi= 1$, the de Donder gauge.} given in Ref.\cite{Han-Lykken} 
are presented in the Appendix. 

After adding coherently these new amplitudes to the 
standard model ones, we get the total amplitude. The new amplitude 
now depends on Higgs boson mass($M_H$), center-of-mass energy($\sqrt{s}$), 
fundamental Planck scale ($M_S$), and finally on the number of extra dimensions
$(n$). Through out our analysis we have fixed the standard model Higgs boson 
mass at a representative value $M_H = 120 $ GeV. 
For center of mass energy, we consider three possible values 
of $\sqrt{s}= 0.5$ TeV, $1$ TeV and $3$ TeV at which the future 
$e^+e^-$ colliders are expected to operate~\cite{TESLA,ANLC,CLIC}.  

In Figure 2, we represent the total cross-section for $M_H= 120 $ GeV and 
$n=3 $ as a function of machine energy for three different values of 
$M_S = 3.5, 4.5$ and $5.5$ TeV. The dashed line represents
the standard model prediction and this exhibits the expected fall-off 
behavior with energy. For large values of $M_S$, this behavior 
is preserved, since the graviton contribution then becomes small. However,
when $M_S$ is smaller, the cross-section show a marked increase with 
energy, showing the well-known feature of gravitational interaction.  
It is obvious from this Figure, that at energies around 3 TeV, the 
graviton contribution is enormous, if $M_S$ is small
( 3.5 TeV); however, a discernible difference exists even when 
$M_S = 5.5 $ TeV. Thus we can expect larger effects, or in other words 
stronger bounds on $M_S$ as the machine energy increases. 

In Figure 3, we represent the total cross-section as a function of $M_S$ for
different values of extra dimensions $n=3,4,5,6$ for two value of machine
energies, $\sqrt{s}=500 $~GeV and 1 TeV. For comparison the standard model
cross-sections are given by the dashed lines, which are  
independent of $M_S$. From this Figure, we can see that for a given 
machine energy and $M_S$, the graviton contribution decreases with the
increase in extra-dimension $(n)$. For a given $n$, the graviton 
contribution asymptotically tends to the standard model value with 
increase of $M_S $ as it should. 

In Figure 4, we represent the angular distribution of the Higgs boson. 
The standard model distribution (dashed line) shows typical 
$\sin^2\theta $ behavior, whereas, the ADD contribution (solid line),
which has in addition spin-2 and spin-0 exchange, can be distinguished from the
standard model very easily. From this distribution, it is evident that 
larger deviation from the standard model is expected in the   
forward and backward regions of the detector, 
with $0.5 < \mid \cos\theta\mid < 0.8 $. We assume that the efficiency of the
detectors in that region will be good enough to observe such events.    
In the central region ( $\mid \cos\theta \mid < 0.2 $ ) of the
detector, the standard model cross-section shows a peak over the 
ADD contribution.  

In Figure 5, we represent the fractional deviation 
$R= \frac{\mid \sigma(M_S) - \sigma_{SM}\mid}{\sigma_{SM}}$ as a function 
of $M_S$ for two values of center of energies, 500 GeV (dashed line), 
and 1 TeV ( solid line ), where, $\sigma(M_S)$ is obtained from coherent
sum of ADD and standard model amplitudes.  
The different choices of extra dimensions $(n)$ are shown along each 
curve. It is very clear from this Figure, that higher the machine energy,
larger is the deviation from the standard model. 
We consider that more than 
$10\%$ deviation $(R = 0.1)$ from the standard model 
prediction with an integrated luminosity ${\cal L}$ of $2{\rm ab}^{-1}$ 
will be able to provide reasonable evidence of low-scale gravity. This 
argument is based on the result obtained in Ref.\cite{zhh_sim}. In this paper,
the authors have studied the standard model $ZHH$ production at future 
$e^+e^-$ linear collider. They have shown that at $\sqrt{s}= 500$ GeV, the
$e^+e^- \to ZHH$ cross-section can be determined with a 
$10\%$ relative error
assuming an integrated luminosity of $2 {\rm ab}^{-1}$. 
In our analysis we use this error ($10\%$) on the cross-section measurement to 
determine $5\sigma $ discovery limit on $M_{S}$ with the same integrated 
luminosity. 

For higher center-of-mass energies$(\sqrt{s}= 1$ TeV and 3 TeV), we 
also assume for simplicity the same relative error in the $ZHH$ cross-section
measurement with the same integrated luminosity. 
We think as a first exercise of this kind, the 
assumption considered here is sufficient. Before we display our 
limits on the scale $M_S$, we would also 
like to stress that the limits on the string scale $M_S$ obtained
here are merely indicative. These limits may change when one considers
detector simulations with full estimations of signal and backgrounds.
As we have already mentioned, this process should not be considered as
the main channel to probe the low scale quantum gravity. Since, by the time
the next future $e^+e^-$ collider starts operating and reaches such a high 
integrated luminosity, the low scale quantum gravity model 
would have been discovered if true. 
 
In  the Table 1, 
we display the $5\sigma $ discovery limits on $M_S$ 
 as a function of $e^+ e^-$ center 
of mass energy $(\sqrt{s})$, number of extra-dimensions $(n)$ and 
integrated luminosity. 
From this Table, we can appreciate the following features:
\begin{itemize}
\item { The $5\sigma $ limit on $M_S$ is weekly dependent on number of
  extra dimensions $(n)$. }
\item { The $5\sigma $ limit on $M_S$ obtained for 500 GeV machine is not very 
promising. In this case the strongest limit is less than 1 TeV for $n=3$. }
\item {The situation become better once we go to a higher center of mass 
energy. The strongest limit on $M_S$ is 6.6 TeV obtained 
for $n=3 $ and at $\sqrt{s}=3 $ TeV.}  

\end{itemize}

\begin{table}
\begin{center}
\begin{tabular}{|c|c|c|c|c|}
\hline
&\multicolumn{4}{|c|}{ }\\
$\sqrt{s}$&\multicolumn{4}{|c|}{$M_S$ in GeV}\\[3mm]
\cline{2-5}
&&&&\\
TeV&$n=3$&4&5&6\\[3mm]
\hline
&&&&\\
0.5  & 921 & 769& 697 & 653 \\
&&&&\\
1.0&1998&1811& 1685 & 1593\\
&&&&\\
3.0&6598 &6036 & 5634 & 5332 \\[3mm]
\hline
\end{tabular}
\caption{$5\sigma $ limit on $M_S$ assuming an integrated 
luminosity of $2~{\rm ab}^{-1}$ for different center of mass 
energies ($\sqrt{s}$). $n$ is the number of extra dimensions.}
\end{center}
\end{table}
  
\section{$e^+e^-\to HH \gamma $}

For the completeness of our study, in this section we discuss the 
production of $HH\gamma$ at $e^+e^-$ collider
through the virtual exchange of spin-2 and spin-0 KK gravitons. The 
scalar pair production at $e^+ e^-$ collider in the models of large
extra-dimensions has been studied~\cite{rizzo_Higgs}. The Feynman diagrams
for $e^+ e^- \to HH\gamma $ are similar to Figure 1, with $Z$ replaced by 
$\gamma $. 
In the case of exchange of spin-2 gravitons all the four diagrams $(a-d)$ 
contribute, while for spin-0 KK mode exchange, only diagrams $(a)$,$(b)$ 
and $(d)$ contribute. These three diagrams  form a gauge invariant set, 
while the diagram $(c)$ vanishes due to Dirac equation and gauge invariance.
In this process, unlike Higgs boson pair production, spin-0 mode 
of KK graviton also contribute. The photon will serve as an additional
trigger. We impose the following identification criteria for the photon 
~\cite{gunion}:
\bea 
E_{\gamma} & > & 10~~{\rm GeV}\\
\mid \cos\theta_{\gamma} \mid & < & 0.98 
\eea
where, $\theta_{\gamma}$ defines the photon angle with the $e^-$ beam 
direction. This angular cut on the photon removes the collinear 
divergence. These `acceptance' cuts are more-or-less the basic ones. 
Though further
selection cuts will become appropriate when a more detailed analysis is done,
it suffices for our analysis, which is a preliminary study. 

Unlike the previous case, here the whole contribution comes
from the exchange of virtual gravitons. As before, we fix the 
Higgs boson mass at the value used in the last section as also the 
machine energies.

In Figure 6, we represent the total cross-section for $M_H= 120 $ GeV and 
$n=3 $ as a function of machine energy for three different values of
$M_S = 3.5, 4.5$ and $5.5$ TeV. 
For a given $M_S$, the cross-section shows a marked increase with 
energy, depicting the well-known feature of gravitational interaction.  

It is obvious from this Figure, that at energies around 3 TeV, the 
graviton contribution is enormous, if the $M_S$ is small
( 3.5 TeV); however, a discernible difference exists even when 
$M_S = 5.5 $ TeV. Thus we can expect larger effects, or in other words 
stronger bounds on $M_S$ as the machine energy increases.  
In Figure 7, we represent the angular distribution of the Higgs boson. 
This particular shape of the distribution is specific to virtual 
exchange of gravitons in the process.

In Figure 8, we represent the total production cross-section as a function
of $M_S$ for different values of  extra-dimensions $n=3,4,5,6$ for three 
values of center-of-mass energies, $\sqrt{s}= 500 $ GeV (dotted lines), 
1 TeV (dashed lines ) and 3 TeV ( solid lines). For a given $\sqrt{s}$ and
$M_{S}$, the graviton contribution decreases very slowly with increase of 
extra-dimension $n$.

The main possible source of background 
to this process is $e^+e^- \to ZH \gamma$ with $Z$ decays to $b\bar b $ looking
like a Higgs decay. Using the CompHEP~\cite{comphep} we have computed
$\sigma^B_{eff}(= \sigma (e^+e^-\to ZH\gamma)\times Br(Z\to b {\bar b}) )
\times Br (H \to b {\bar b}) )$ 
for $e^+ e^- \to ZH\gamma $ process for three values of 
$\sqrt{s}=0.5$ TeV,~1~TeV and 3 TeV to get a feeling for this 
background. The numbers are 0.50 fb,~0.15 fb and 0.023 fb for 
$\sqrt{s}=0.5$ TeV, 1~TeV and 3 TeV respectively. 
There are two ways one can 
eliminate this background. Firstly, if the Higgs mass is heavier than $M_Z$
(which is in fact true from LEP II data)~\cite{LEP2_Higgs}, the invariant mass 
distribution of particles produced from Higgs decay will be able 
to reduce the background. Secondly, it should also be possible to reduce the 
background further by studying the angular distribution associated with 
these processes. The loop-induced standard model or MSSM Higgs pair production 
cross-section for light Higgs mass is of the order $0.1-0.2 $ fb at 
$\sqrt{s}= 500 $ GeV~\cite{loop_hpair}. This cross-section will be 
suppressed by order 
$\alpha_{em} $, if an additional photon is produced. When folded
with $H\to b \bar b $ branching ratio, the cross-section becomes 
$O(10^{-4})$ fb. This cross-section will be further suppressed 
by the 3 body phase-space.    

To estimate the possible $5\sigma$ discovery limit on the scale $M_S$
with an integrated luminosity of $2 {\rm ab}^{-1}$, we only consider 
the $ZH\gamma$ as the main source of standard model background. We multiply the standard
model cross-section by the luminosity to get the predicted number of events. 
We then estimate the errors assuming that the statistical errors are Gaussian
and that there are no systematic errors. This certainly makes our
estimates of the discovery limits over-optimistic ( especially for 
for $\sqrt{s}= 3 $ TeV, where number of standard model events are small ). 
In any case, before more detailed studies of the standard model backgrounds
and the detector design, any estimate of errors must be considered a 
crude estimate. 
In Table~2 we display such discovery limits on $M_S$ for different 
machine energies and extra-dimensions. As expected the strongest limit 
appears for $\sqrt{s} =3 $ TeV and for $n=3$.

\begin{table}
\begin{center}
\begin{tabular}{|c|c|c|c|c|}
\hline
&\multicolumn{4}{|c|}{ }\\
$\sqrt{s}$&\multicolumn{4}{|c|}{$M_S$ in GeV}\\[3mm]
\cline{2-5}
&&&&\\
TeV&$n=3$&4&5&6\\[3mm]
\hline
&&&&\\
0.5  & 1557 & 1175 & 993 & 883 \\
&&&&\\
1.0&2847 &2154 & 1834 & 1637\\
&&&&\\
3.0&7383 & 5673 &4866 &4383 \\[3mm]
\hline
\end{tabular}
\caption{$5\sigma $ discovery limit on $M_S$ assuming an integrated luminosity
of $2~{\rm ab}^{-1}$ for different center-of-mass energies 
($\sqrt{s}$). $n$ is the number of extra dimensions. }
\end{center}
\end{table}

\section{Conclusions}

In this paper we have studied the implications of KK graviton contribution
to the process $e^+e^- \to HHZ$, which has been studied in the standard 
model. The spin-2 and spin-0 mode of KK gravitons contribute to this
process substantially. However, the existence of low-energy quantum 
gravity may be discovered through more direct channels as studied by
several groups. Nevertheless, this process will be an independent 
confirmation for such a discovery. The $5\sigma$ discovery 
limits on the string scale $M_S $ is obtained assuming an integrated 
luminosity of $2~{\rm ab}^{-1}$. We have derived this limit considering
$10\%$ error on the $HHZ$ cross-section measurement at $\sqrt{s}= 500 $
GeV at ${\cal L}= 2~{\rm ab}^{-1}$. In reality for higher center-of-mass 
energies, the above error may change. However, in our analysis we have assumed
it to be remain the same. The bounds obtained here are only suggestive and 
may change when one
take into account proper detector simulation, inclusing 
$H$ and $Z$ decays and imposing selection cuts on the final state particles
and the full background calculation.  

Another interesting behavior is shown by the Higgs angular distributions. 
In the standard model, the 
shape behaves like $\sim \sin^2\theta $, whereas it is completely different
in the presence of KK gravitons. Furthermore, we have shown that in the
forward and backward regions of the beam pipe, one would expect larger
deviation from the standard model prediction. Hence, careful study of Higgs
angular distribution may provide evidence of new physics beyond the 
standard model. 

To complete our analysis we have also studied the $e^+e^- \to HH\gamma$ 
process in this model which is absent at the
tree level in the standard model.
We have discussed how this mode can be distinguished 
from the possible backgrounds. 
We have obtained the $5\sigma $ discovery limit on $M_S$ assuming 
an integrated luminosity of $2~{\rm ab}^{-1}$. 

\section{Appendix}
Here we will write down the graviton contribution ( both spin 2 and $0$ ) 
amplitudes for the process :$e^+e^- \to HHZ$. In our analysis we have 
chosen $\xi = 1$, the so called de Donder gauge. The Feynman rules 
used to obtain these matrix elements have been taken from 
Ref.\cite{Han-Lykken}. The relation between the fundamental scale $M_S$ and
the size $R$ of the $n$ extra dimensions is given by \cite{Han-Lykken}
\bea
\kappa^2 R^n = 16 \pi (4\pi)^{n/2} \Gamma(n/2) M_S^{-(n+2)}
\eea
where, $\kappa = \sqrt{16 \pi G_N}$. $G_N $ is the four dimensional Newton 
constant. 
The amplitudes $a-d$ and $e-h$ correspond to the exchange of a virtual 
graviton of spin $2$ and $0$ respectively.
\footnotesize
\bea
{\cal M}_{a} &=& - \frac{\pi g}{2 C_W}{\cal G}\frac{P^{\mu\nu\alpha\beta}}
{(k_1 - p_1)^2}{\cal X}_{\alpha\beta} 
\big [ {\bar v(k_2)}C_{\mu\nu}({k_1\hspace{-1em}/\;\:} 
-{p_1\hspace{-1em}/\;\:})\gamma^{\rho}(g_v + g_a \gamma_5 ) u(k_1) \big ] 
\epsilon^{*}_{\rho}(p_1)\\
{\cal M}_{b} &=& \frac{\pi g}{2 C_W}{\cal G}\frac{P^{\mu\nu\alpha\beta}}
{(k_2 - p_1)^2}{\cal X}_{\alpha\beta}
\big [ {\bar v(k_2)}\gamma^{\rho} (g_v+g_a \gamma_5) ({k_2\hspace{-1em}/\;\:} 
-{p_1\hspace{-1em}/\;\:}) C^\prime_{\mu\nu} u(k_1) \big ] \epsilon^{*}_{\rho}(p_1)\\
{\cal M}_{c} &=& \frac{2\pi g}{C_W}{\cal G}\frac{P^{\rho\sigma\alpha\beta}}
{(s - M^2_Z)} {\cal F}_{\rho\sigma\nu\lambda}(g^{\mu\nu}- \frac{q^{\mu}q^{\nu}}
{M^2_Z}){\cal X}_{\alpha\beta}
\big [ {\bar v(k_2)}\gamma_{\mu} (g_v+g_a \gamma_5) u(k_1) \big ] 
\epsilon^{\lambda *}(p_1)\\
{\cal M}_{d} &=& \frac{\pi g}{C_W} {\cal G} P^{\mu\nu\alpha\beta} {\cal X}_{\alpha\beta}
\big ( C_{\mu\nu,\rho\sigma} - \eta_{\mu\nu}\eta_{\rho\sigma} \big )
\big [ {\bar v(k_2)}\gamma^{\sigma} (g_v+g_a \gamma_5) u(k_1) \big ] 
\epsilon^{\rho *}(p_1)\\
{\cal M}_{e} &=& \frac{16\pi g}{C_W}\big ( \frac{n}{-n^2+4}\big ) {\cal G} 
\big ( p_2.p_3 + 2 M^2_H ) 
\big [ {\bar v(k_2) }\gamma^\alpha (g_v + g_a \gamma_5 ) u (k_1) \big ] 
\epsilon^{*}_{\alpha}(p_1)   \\
{\cal M}_{f} &=& {\cal M}_{e}\\
{\cal M}_{g} &=& -\big ( \frac{64 \pi g }{3 C_W}\big ) {\cal G} 
\frac{n~\epsilon^{\mu *}(p_1)}{(-n^2+4) (s - M_Z^2)} \big ( p_2.p_3 + 2 M_H^2 \big ) {\cal Z}_{\beta\mu}
\big [ {\bar v(k_2)}\gamma_{\alpha}(g_v + g_a \gamma_5 ) u (k_1)\big ]
\big ( g^{\alpha\beta} - \frac{q^{\alpha} q^{\beta}}{M_Z^2} \big ) \\
{\cal M}_{h} &=& \frac{32\pi g}{C_W} \big ( \frac{n}{-n^2+4}\big ) {\cal G} 
\big ( p_2.p_3 + 2 M^2_H \big ) 
\big [ {\bar v(k_2)}\gamma^\alpha (g_v + g_a \gamma_5 ) u(k_1) \big ] 
\epsilon^{*}_{\alpha} (p_1) 
\eea
\normalsize
where, $q= (k_1+k_2)$, $k_1$ and $k_2$ are two incoming four momenta. 
$C_W= \cos\theta_W$, $\theta_W$ the Weinberg angle, $g_v$ and
$g_a$ are the vector and axial vector couplings of $Z$ to electron.  
\bea
{\cal G} &=& \frac{\kappa^2}{16\pi} D(s_1)\\
{\cal X}_{\alpha\beta} &=& (M^2_H \eta_{\alpha\beta} 
-C_{\alpha\beta\delta\eta} p^{\delta}_{2}p^{\eta}_{3})\\
{\cal Z}_{\beta\mu} & = & \big ( \eta_{\beta\mu} M_Z^2 + q^\prime_{\beta} q^\prime_{\mu}
-p_{1\mu}q^\prime_{\beta} \big ) \\
C_{\mu\nu} &=& \big [\gamma_{\mu}\big (k_1 - p_1 - k_2)_{\nu} + (\mu \leftrightarrow \nu) \big ] - 2 \eta_{\mu\nu} ({k_1\hspace{-1em}/\;\:} -{p_1\hspace{-1em}/\;\:}-{k_2\hspace{-1em}/\;\:}) \\
C^{\prime}_{\mu\nu} &=& C_{\mu\nu}(p_1 \go -p_1)\\
{\cal F}_{\rho\sigma\nu\lambda} &=& \big [ \big ( M^2_Z - q.p_1 \big ) 
C_{\rho\sigma, \nu\lambda} + D_{\rho\sigma, \nu\lambda}(q, -p_1) 
+ E_{\rho\sigma, \nu\lambda}(q, -p_1) \big ]\\
P^{\mu\nu\alpha\beta} & = & \eta^{\mu\alpha}\eta^{\nu\beta}
                            + \eta^{\mu\beta}\eta^{\nu\alpha}
                            - \frac{2}{(n-2)}\eta^{\mu\nu}\eta^{\alpha\beta}~
({\rm in~de~Donder~gauge}) 
\eea
where, $s_1 = (p_2 + p_3)^2 $, $p_2$ and $p_3$ are four momenta of outgoing
Higgs particles. The function $D(s_1)$ counts for the 
virtual KK state exchanges. Complete expression of $D(s_1)$ can be 
found in Ref.~\cite{Han-Lykken}. $q^\prime = \sqrt{s_1}$ and 
$\eta^{\mu\nu}$ is the Minkowski metric tensor.
The definitions of $C_{\rho\sigma, \nu\lambda}, 
D_{\rho\sigma, \nu\lambda}(q, -p_1)$ and $E_{\rho\sigma, \nu\lambda}(q, -p_1)$
can be found in \cite{Han-Lykken}. The expression for 
$P^{\mu\nu\alpha\beta}$ is
taken from Ref.\cite{rattazzi}. The amplitudes for the process 
$e^+e^- \to HH\gamma $ can be obtained from Equations (5-12), by putting
$M_Z = 0$, $g_v= 1$ and $g_a = 0 $. For spin-2 graviton mode exchange, 
all the four amplitudes in Equations (5-8) will contribute, while for spin-0 
mode, Equations (9,10) and Equation (12) will contribute.

\section{Acknowledgments}
This work was supported by US DOE contract numbers DE-FG03-96ER40969. 
Authors would like to thank M. Perelstein and David Strom for discussions.  

\def\pr#1,#2 #3 { {Phys.~Rev.}        ~{\bf #1},  #2 (19#3) }
\def\prd#1,#2 #3{ { Phys.~Rev.}       ~{D \bf #1}, #2 (19#3) }
\def\pprd#1,#2 #3{ { Phys.~Rev.}      ~{D \bf #1}, #2 (20#3) }
\def\prl#1,#2 #3{ { Phys.~Rev.~Lett.}  ~{\bf #1},  #2 (19#3) }
\def\pprl#1,#2 #3{ {Phys. Rev. Lett.}   {\bf #1},  #2 (20#3)}
\def\plb#1,#2 #3{ { Phys.~Lett.}       ~{\bf B#1}, #2 (19#3) }
\def\pplb#1,#2 #3{ {Phys. Lett.}        {\bf B#1}, #2 (20#3)}
\def\npb#1,#2 #3{ { Nucl.~Phys.}       ~{\bf B#1}, #2 (19#3) }
\def\pnpb#1,#2 #3{ {Nucl. Phys.}        {\bf B#1}, #2 (20#3)}
\def\prp#1,#2 #3{ { Phys.~Rep.}       ~{\bf #1},  #2 (19#3) }
\def\zpc#1,#2 #3{ { Z.~Phys.}          ~{\bf C#1}, #2 (19#3) }
\def\epj#1,#2 #3{ { Eur.~Phys.~J.}     ~{\bf C#1}, #2 (19#3) }
\def\mpl#1,#2 #3{ { Mod.~Phys.~Lett.}  ~{\bf A#1}, #2 (19#3) }
\def\ijmp#1,#2 #3{{ Int.~J.~Mod.~Phys.}~{\bf A#1}, #2 (19#3) }
\def\ptp#1,#2 #3{ { Prog.~Theor.~Phys.}~{\bf #1},  #2 (19#3) }
\def\jhep#1, #2 #3{ {J. High Energy Phys.} {\bf #1}, #2 (19#3)}
\def\pjhep#1, #2 #3{ {J. High Energy Phys.} {\bf #1}, #2 (20#3)}

\begin{figure}[hbt]
\centerline{\epsfig{file=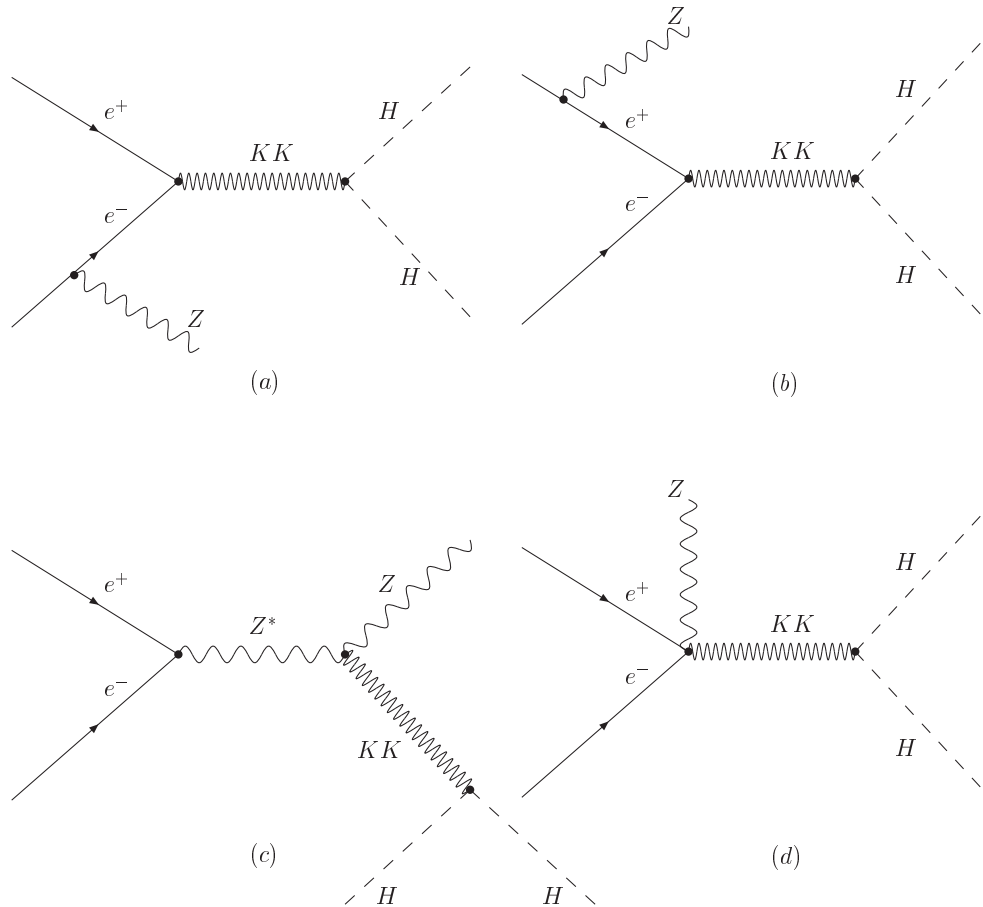,width=\linewidth}}
\vspace*{-8.0cm}
\caption{ Feynman diagrams of virtual gravitons ( both spin-0 and spin-2)
contribution to the process $e^+e^- \to HHZ$.}
\label{fig0}
\end{figure}
\begin{figure}[hbt]
\centerline{\epsfig{file=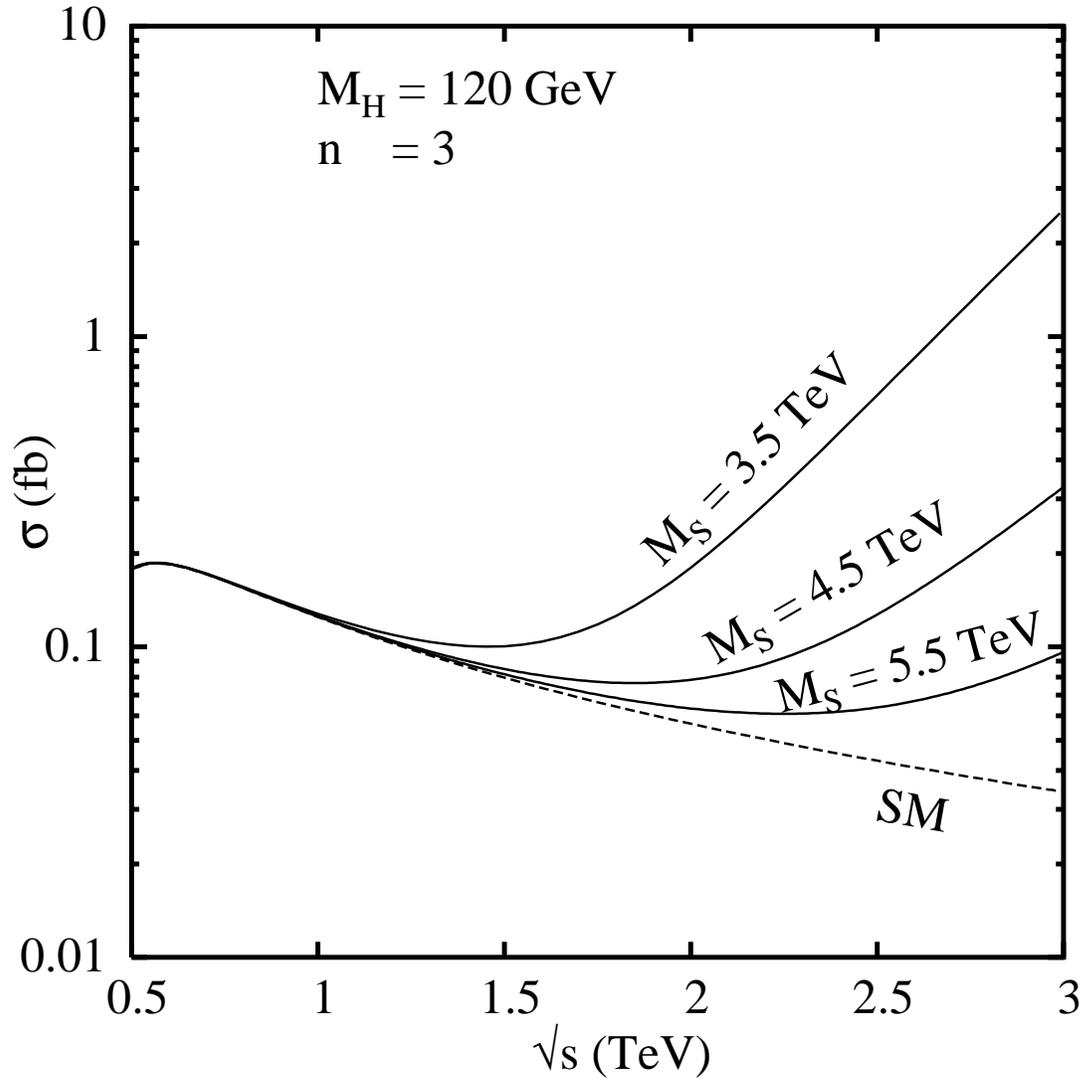,width=\linewidth}}
\vspace*{-8.0cm}
\caption{Variation of the cross-section (fb) for $HHZ$ production with 
machine energy. The solid curves correspond to the ADD predictions
for $M_S = 3.5, 4.5$ and $5.5 $ TeV and $n=3$, while the dashed line represents
only the standard model contributions. The Higgs mass is set to 120 GeV.}
\label{fig1}
\end{figure}

 \begin{figure}[hbt]
 \centerline{\epsfig{file=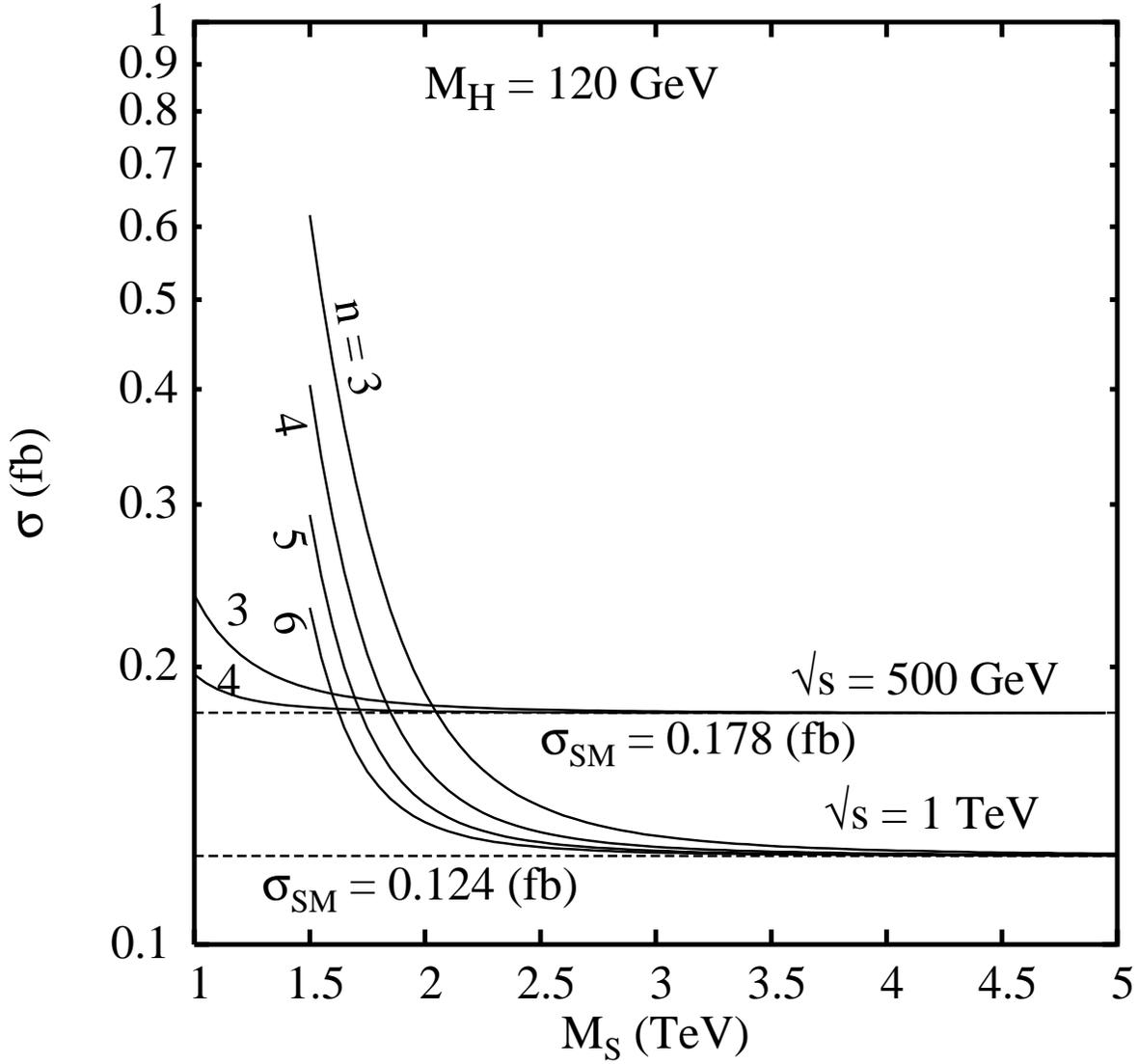,width=\linewidth}}
 \vspace*{-6.0cm}
 \caption{Variation of the cross-section (fb) for $HHZ$ production 
 with $M_S$ for $\sqrt{s} = 0.5 $ and 1 TeV are shown by the solid lines. 
 The values of extra dimensions ($n$) are given along the each solid 
 curves. The corresponding standard model contribution is shown by the dashed line with
 its value shown along it. The Higgs mass is same as Figure 1. }
\label{fig2}
\end{figure}

 \begin{figure}[hbt]
 \centerline{\epsfig{file=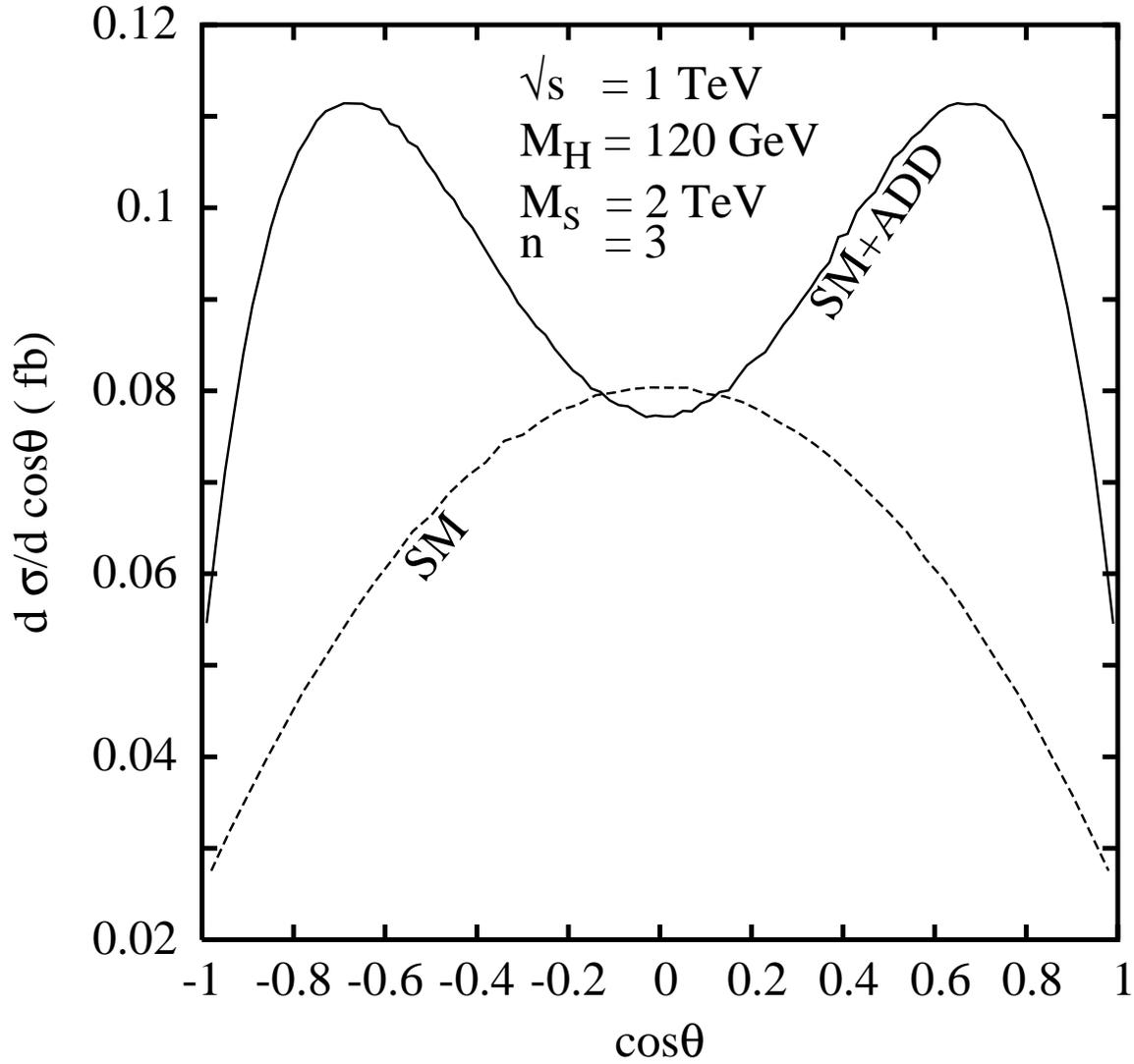,width=\linewidth}}
 \vspace*{-6.0cm}
 \caption{ The tree level differential cross-section for $e^+e^- \to ZHH$
 at 1 TeV $e^+e^-$ linear collider. The solid line correspond to ADD  predictions for $M_S = 2 $ TeV and $n = 3$, while the dashed line 
 represent the standard model contribution, assuming $M_H = 120 $ GeV. }
\label{distfig}
\end{figure}
 \begin{figure}[hbt]
 \centerline{\epsfig{file=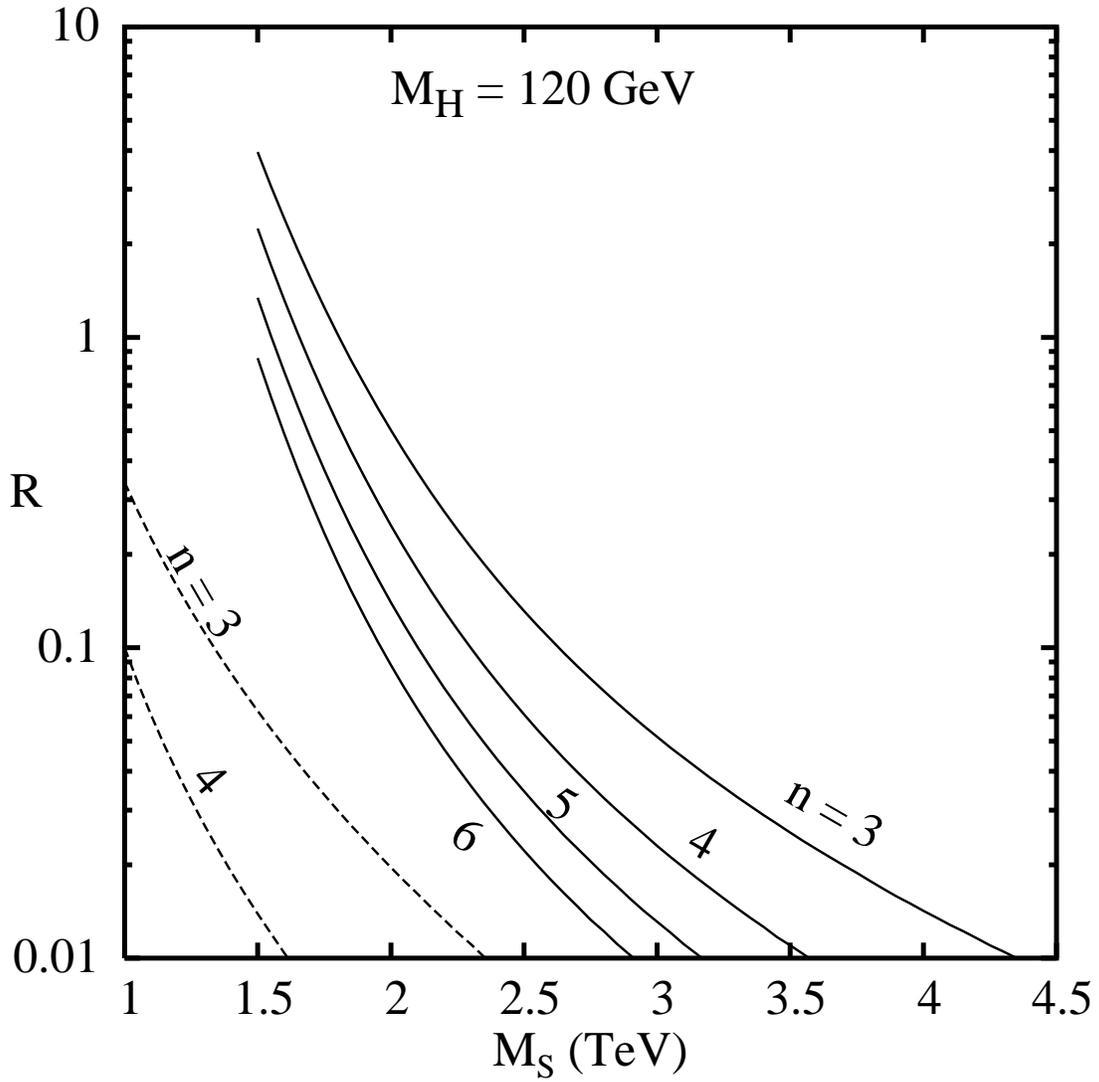,width=\linewidth}}
 \vspace*{-6.0cm}
 \caption{The fractional deviation $R$( defined in the text) 
 of the cross-section for $e^+e^- \to HHZ $
 from the standard model prediction as a function of $M_S$ for different choices of extra
 dimensions as shown along the curves. The solid and dashed curves correspond 
 to $\sqrt{s} = 1$ TeV and $0.5$ TeV NLC. The Higgs mass is same as Figure 1.}
\label{fig3}
\end{figure}
\begin{figure}[hbt]
\centerline{\epsfig{file=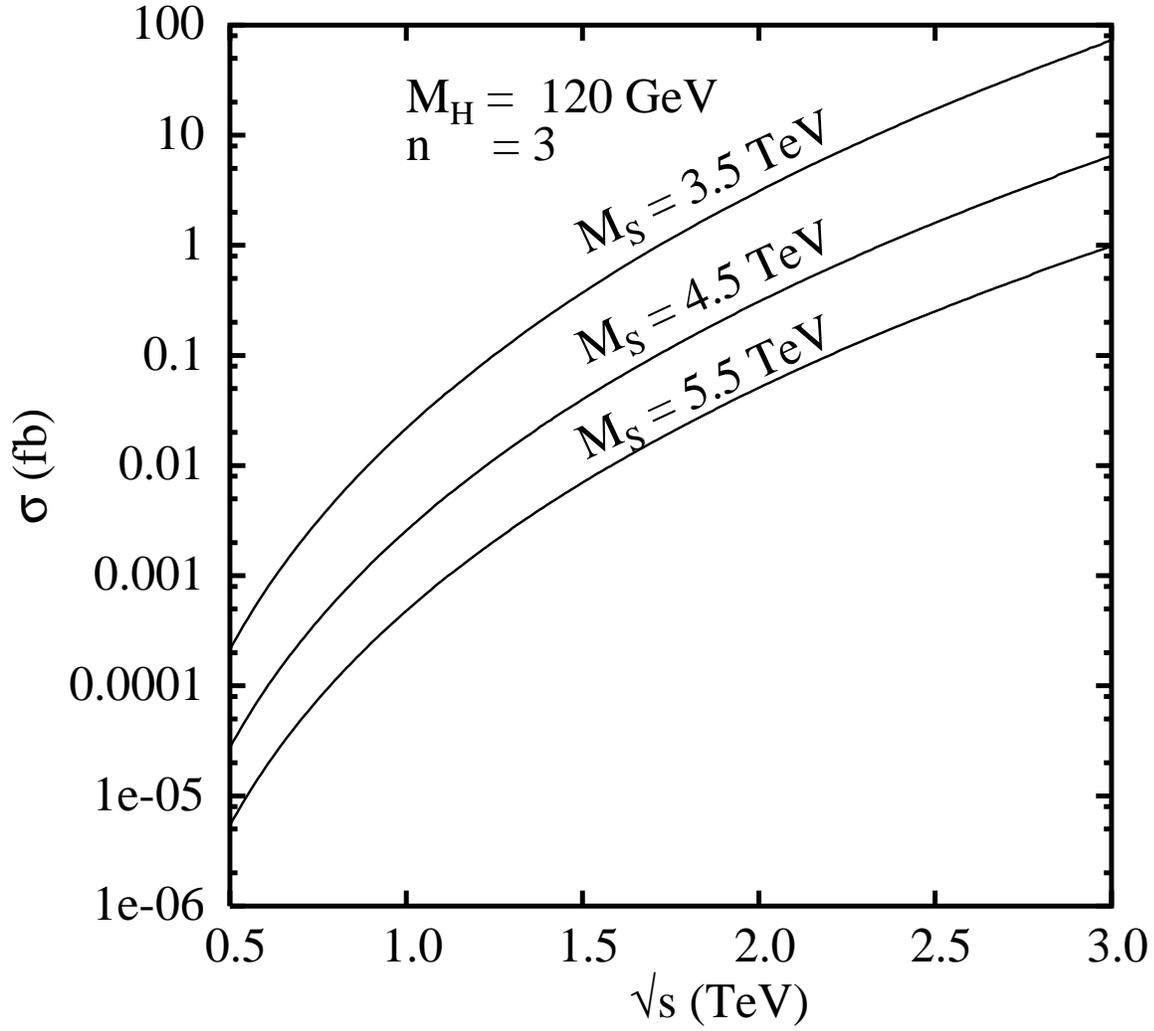,width=\linewidth}}
\vspace*{-8.0cm}
\caption{Variation of the cross-section (fb) for $HH\gamma$ production with 
machine energy in the ADD model with $M_S = 3.5, 4.5$ and $5.5 $ TeV and $n=3$.
The Higgs mass is set to 120 GeV.}
\label{fig4}
\end{figure}

 \begin{figure}[hbt]
 \centerline{\epsfig{file=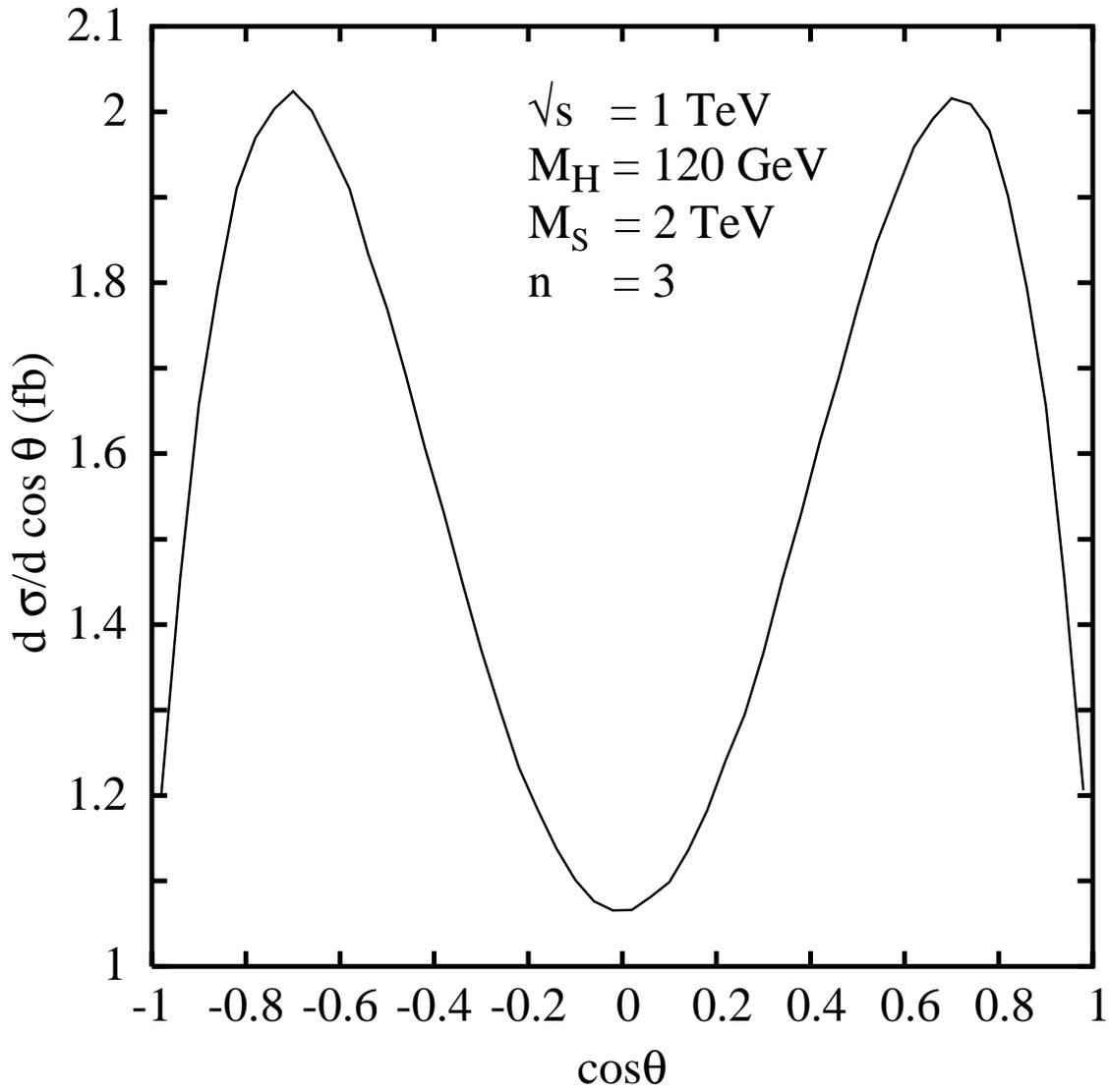,width=\linewidth}}
 \vspace*{-6.0cm}
 \caption{The tree level differential cross-section for $e^+e^- \to HH\gamma$
 at 1 TeV $e^+e^-$ linear collider, $M_S = 2 $ TeV, $n =3 $ and 
$M_H = 120 $ GeV. }
\label{ahh_dist}
\end{figure}

 \begin{figure}[hbt]
 \centerline{\epsfig{file=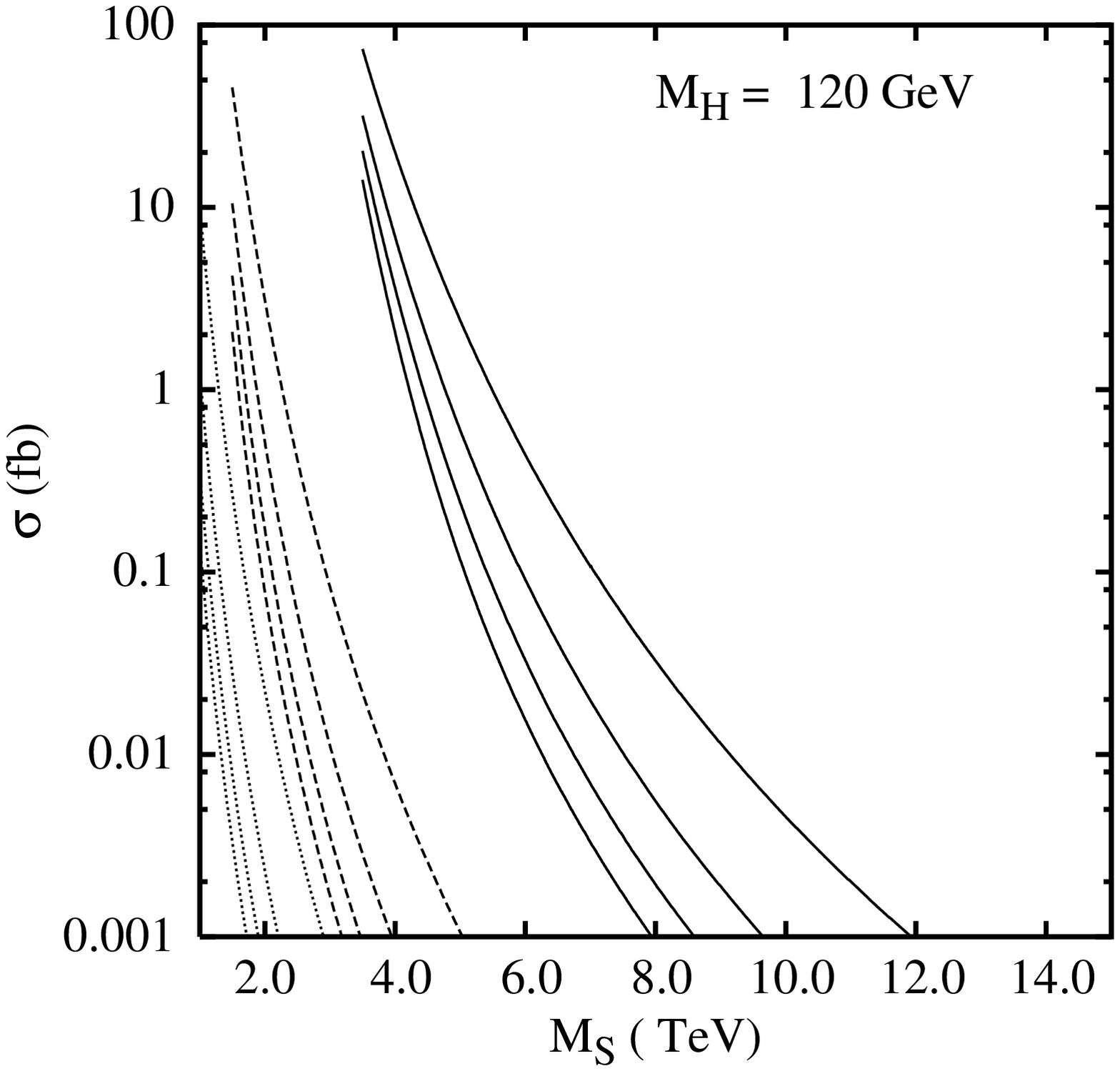,width=\linewidth}}
 \vspace*{-6.0cm}
 \caption{Variation of the cross-section (fb) for $HH\gamma$ production 
 with $M_S$. The solid, dashed and dotted curves with the extra 
 dimension $n=3-6$ from right to left  correspond to 
 $\sqrt{s} =3 $ TeV, 1 TeV and 0.5 TeV respectively.}
\end{figure}


\begin{thebibliography}{99}
\bibitem{ADD}
N.~Arkhani-Hamed, S.~Dimopoulos and G.~R.~Dvali, \plb429,263 {98};
\prd59,086004 {99}; I.~Antoniadis, N.~Arkani-Hamed, S.~Dimopoulos and 
G.~R.~Dvali, \plb436,257 {98}.
\bibitem{GRW}G.~F.~Giudice, R.~Rattazzi, and J.~D.~Wells, \npb544,3 {99};
\bibitem{Han-Lykken}T.~Han, J.~D.~Lykken and Ren-Jie Zhang, \prd59,105006 {99}.
\bibitem{ADD_rev} For reviews see : T.~G.~Rizzo, hep-ph/9911229; 
Yuri A.~Kubyshin, eprint hep-ph/0111027; 
J.~Hewett and M.~Spiropulu, hep-ph/0205106 and references therein.
\bibitem{TESLA} R.~D.~Heuer \etal, TESLA Technical Design Report: Part III, 
DESY-2001-011 (hep-ph/0106315); 
\bibitem{ANLC} G.~Pasztor and T.~G.~Rizzo, Snowmass 2001, hep-ph/0112054
\bibitem{CLIC} The CLIC Study Team, {\it A 3 TeV $e^+e^-$ linear collider 
based on CLIC technology}, CERN 2000-008, D.~Schulte, 
see http://clicphysics.web.cern.ch/CLICphysics/
\bibitem{zerwas} A.~Djouadi, W.~Kilian, M.~Muhlleitner and P.~M.~Zerwas, 
\epj10,27 {99}

\bibitem{zhh_sim}C.~Castanier, P.~Gay, P.~Lutz and J.~Orloff, hep-ex/0101028.
 
\bibitem{rizzo_Higgs} T.~G.~Rizzo, \prd60,075001 {99}.
\bibitem{gunion}J.~F.~Gunion and S.~Mrenna, \pprd64,075002 {01}. 

\bibitem{comphep}A.~Pukhov, \etal, hep-ph/9908288
\bibitem{LEP2_Higgs} The LEP Working Group for Higgs Boson Searches, Search
for the Standard Model Higgs Boson at LEP, LHWG Note/2002-01.

\bibitem{loop_hpair}K.~J.~F.~Gaemers and F.~Hoogeveen, \zpc26,249 {84}; 
A.~Djouadi, V.~Driesen, and C.~Junger, \prd54,759 {96}.
\bibitem{rattazzi}R.~Contino, L.~Pilo, R.~Rattazzi, and A.~Strumia, 
JHEP {\bf 0106}, 005 2001.

\end{thebibliography}
\end{document}